\documentclass[twoside,12pt]{article}
\usepackage[dvips]{epsfig}
\usepackage{graphicx}

\def\p{\vec p}

\newcommand{\be}{\begin{equation}}
\newcommand{\ee}{\end{equation}}
\newcommand{\bea}{\begin{eqnarray}}
\newcommand{\eea}{\end{eqnarray}}

\begin{document}

\title{ \vspace{1cm} Two-flavor QCD at finite temperature 
and chemical potential in a functional approach}
\author{Jan Luecker$^1$, Christian S. Fischer$^{1,2}$  \vspace*{0mm}\\
$^1$ Institut f\"ur Theoretische Physik,
  Justus-Liebig-Universit\"at Gie\ss{}en, \\
  Heinrich-Buff-Ring 16,
  D-35392 Gie\ss{}en, Germany \\
$^2$ GSI Helmholtzzentrum f\"ur Schwerionenforschung GmbH, \\
  Planckstr. 1  D-64291 Darmstadt, Germany \\
}
\maketitle
\begin{abstract}
We summarize recent results obtained in the Dyson-Schwinger formalism to study the chiral 
and deconfinement phase transitions of quenched and unquenched QCD at finite temperature 
and chemical potential.
In the quenched case we compare $SU(2)$ and $SU(3)$ gauge theories by taking lattice data
for the gluon as an input for the quark Dyson-Schwinger equation. As compared to previous
investigations we find a clearer distinction between the second order transition of the 
two-color theory and the (weak) first order transition of the three-color gauge theory. 
We then extend this study to unquenched QCD at finite chemical potential by taking matter 
effects to the gluon into account and investigate the order of the chiral phase transition 
and the behavior of the deconfinement transition.
What we find are coinciding phase transitions up to a critical endpoint which is located
at large chemical potential.
\end{abstract}

\section{Introduction}
The behavior of quantum chromodynamics at large temperatures and densities received a 
lot of attention over the past years and is an ongoing research program from both,
theoretical and experimental side. At vanishing chemical potential lattice QCD has 
shown the existence of a crossover from the chiral symmetry broken and confined 
hadronic phase to the phase of the (approximately) chiral symmetric and deconfined 
quark-gluon plasma. While this has been confirmed many times, the behavior
at finite chemical potential is still under intense debate. Lattice methods have a 
limited applicability here due to the fermion sign problem.

Models like the Nambu--Jona-Lasinio (NJL) and the quark-meson (QM) model have so far been the
main source of studies at large chemical potential, and established a scenario where
the chiral crossover turns into a first order phase transition at a critical
endpoint. In Polyakov loop extended versions these models (PNJL and PQM) have also been used to 
study the confinement/deconfinement transition, which may or may not coincide with 
the chiral transition \cite{Fukushima:2003fw,Schaefer:2007pw,Herbst:2010rf,Skokov:2010uh}.
At large chemical potentials and relatively small temperatures there may also
exist some new phases, e.g. color superconductors, inhomogeneous 
\cite{Kojo:2009ha,Nickel:2009ke} or quarkyonic \cite{McLerran:2007qj} phases.

Another direct approach to non-perturbative QCD without the sign problem is the
framework of Dyson-Schwinger equations (DSEs) \cite{Fischer:2009wc,Fischer:2010fx,Fischer:2011mz}
and the functional renormalisation group \cite{Braun:2007bx,Braun:2009gm}.
QCD with two degenerate quark flavors has been studies in Ref.~\cite{Fischer:2011mz}
by solving the coupled system of quark and gluon DSEs using quenched lattice
data for the gluon propagator as input. Within this truncation scheme the behavior 
of the chiral and deconfinement transitions at finite chemical potential have been 
investigated using the first calculation of the dressed Polyakov loop in this region 
of the QCD phase diagram. In this proceedings contribution we give an overview of the
employed truncation scheme and summarize the corresponding results.

\section{Order parameters for chiral symmetry breaking and confinement}
The central object of our investigations is the in-medium quark propagator
which can be decomposed as

\be
S^{-1}(p) = i\vec{\gamma}\vec{p}A(\omega_n,\p^2)+i\gamma_{4}(\omega_n+i\mu)C(\omega_n,\p^2) + B(\omega_n,\p^2),
\ee
where $\mu$ is the quark chemical potential and $\omega_n = \pi T (2n+1)$ 
are the Matsubara modes in the imaginary time formalism with temperature $T$.
The functions $A$, $C$ and $B$ dress the vector and scalar part of the propagator
which we calculate from the corresponding DSE.
The bare propagator $S_0$ at quark mass $m$ is characterized by $A=C=1$ and $B=m$.
A quark propagator with a non-vanishing $B$ function corresponds to a phase
of broken chiral symmetry, either explicitly by a bare quark mass or dynamically
by quantum effects.

A possible order parameter for chiral symmetry breaking is the quark condensate

\be
\langle\bar\psi\psi\rangle = \mathrm{Tr}[S]
= Z_2 Z_m T\sum_n \int\frac{d^3 p}{(2\pi)^3} \frac{4\cdot B(\omega_n,\p^2)}{\p^2 A^2(\omega_n,\p^2) + \omega_n^2 C^2(\omega_n,\p^2) + B^2(\omega_n,\p^2)},
\ee
which can be calculated directly from a solution of the quark DSE.
The condensate in this definition is divergent with $m\Lambda^2$ and $m^2\Lambda$,
but since these terms do not depend on temperature and chemical potential the condensate
can still be used as an order parameter.
We work with approximately physical quark masses and therefore expect to find a crossover
at small chemical potentials. To define the pseudo-critical temperature in this case
we use the susceptibility

\be
\chi=\frac{\partial \langle\bar\psi\psi\rangle}{\partial m},
\ee
and determine its maximum to find $T_c$. The divergent terms in the condensate only
lead to an offset in $\chi$, without changing its maximum.
The quark condensate in the chiral limit has been determined in
Ref.~\cite{Fischer:2011pk}, where critical scaling beyond the mean field level 
at the second order chiral phase transition has been studied.

Constructing an order parameter for confinement that is accessible with functional methods 
is a more challenging task. In the quenched case the Polyakov loop is an accepted order 
parameter which is known to show a (weak) first order phase transition for QCD at a temperature 
of approximately $270$ MeV and a second order transition for the two-color gauge theory
at approximately $300$ MeV.
The Polyakov-loop expectation value is sensitive to breaking of centre symmetry,
which occurs in the deconfined phase, while the confined phase is centre symmetric.
However, if finite quark masses are taken into account within QCD, centre symmetry is broken
explicitly (just like chiral symmetry). There are, however, interesting indications 
that this might not be the case if the larger centre symmetry of the full standard 
model is taken care of \cite{Edwards:2010ew}.

In \cite{Gattringer:2006ci,Synatschke:2007bz,Bilgici:2008qy} the so-called dual condensates have been proposed as order parameters for
centre symmetry breaking. They are defined as

\be
\Sigma_n = \int_0^{2\pi}\frac{d\varphi}{2\pi}e^{-i\varphi n} \langle\bar\psi\psi\rangle_\varphi,
\ee
where $\varphi \in [0,2\pi[$ is a parameter for $U(1)$-valued boundary conditions of
the quark fields: $\psi(\vec{x},1/T)=e^{i\varphi}\psi(\vec{x},0)$. The physical boundary condition is $\varphi=\pi$. The quantity
$\Sigma_n$ contains all loops of connections winding $n$ times around the Euclidean time direction.
For $n=\pm 1$ this also contains the Polyakov loop, and $\Sigma_{\pm 1}$ has therefore been
named the `dressed Polyakov loop'. It also contains all kinds of loops which are not straight in
the time direction but contain detours, but these loops are $1/m$ suppressed. In the $m\rightarrow\infty$
limit the normal Polyakov loop is recovered.
At finite chemical potential $\Sigma_{+1}$ and $\Sigma_{-1}$
are not equal and correspond to the dressed Polyakov loop and its conjugate.
Since the dressed Polyakov loop is sensitive to centre symmetry breaking, it allows us
to calculate the confinement/deconfinement transition from solutions of the quark DSE
without having to deal with an {\it ansatz} for a Polyakov-loop potential as necessary
in the PNJL and PQM models.

Since the deconfinement transition will be a crossover in the unquenched case, we use the maximum of
$\frac{\partial\Sigma_{\pm 1}}{\partial m}$ to define the pseudo-critical temperature,
similar to the chiral transition.

\section{Dyson-Schwinger equations}
\begin{figure}
\centering
\includegraphics[width=0.8\textwidth]{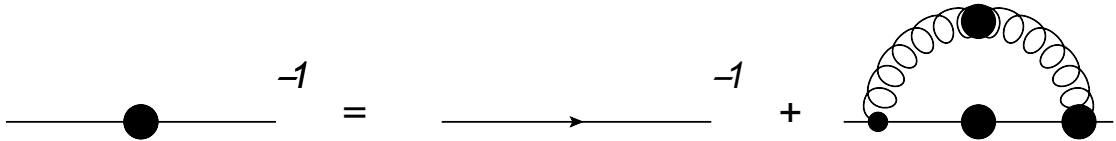}
\caption{The Dyson-Schwinger equation for the quark propagator.
Dots denote dressed objects.}\label{fig:quarkDSE}
\end{figure}
Fig.~\ref{fig:quarkDSE} displays the DSE for the quark propagator.
The quark self-energy depends on the fully dressed gluon and quark-gluon vertex, which
we need to specify in order to solve the equation self-consistently.
For the in-medium gluon propagator we take two steps, first we investigate quenched
QCD where lattice calculations are up to now the most reliable source for the
temperature dependent gluon propagator, and then we will introduce unquenching effects 
by resorting to the gluon DSE.

\begin{figure}
\centering
\includegraphics[width=0.8\textwidth]{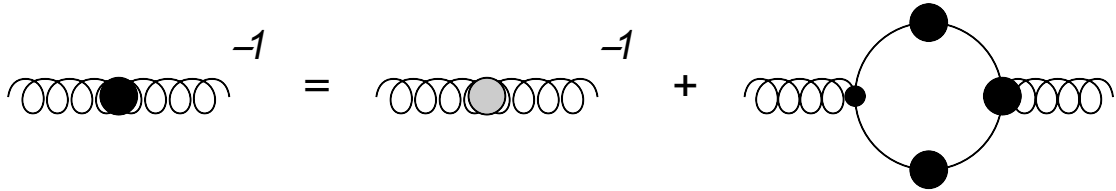}
\caption{The truncated gluon DSE. The black dot denotes the full, unquenched propagator,
while the grey dot denotes the quenched propagator.} \label{fig:gluonDSE}
\end{figure}

The unquenched gluon DSE in the truncation we use here is depicted in Fig.~\ref{fig:gluonDSE}.
The full propagator is given by the lattice results for the quenched propagator
and the quark loop. This is only an approximation, since all diagrams where quark loops
appear inside the Yang-Mills part of the self-energy are neglected. In vacuum
this leads to an over-estimation of the quark loop on the few percent level.
Assuming this still holds at finite temperature, we expect to underestimate the critical
temperatures by up to $10$ MeV.

The quark loop shown in Fig.~\ref{fig:gluonDSE} is given by

\be
\Pi_{\mu\nu}(p) = \frac{Z_{1F} N_f}{2} \sum_n\int\frac{d^3 k}{(2\pi)^3}\,\, \mathrm{Tr}
\left[S(q)g\gamma_{\mu}S(k)g\Gamma_{\nu}\right],\label{eq:quark-loop}
\ee
where $q=p+k$, $Z_{1F}$ is the vertex renormalization factor and $\Gamma_\nu$ the full
quark-gluon vertex.
In principle the system of quark and gluon DSEs can now be solved self-consistently
but as a first approximation we will treat the quark loop semi-perturbatively by taking
bare quarks but a dressed vertex. This allows us to use the hard-thermal loop expression
multiplied by the vertex dressing function, defined below.
The HTL approximation is well justified above the critical temperature where quark dressing
effects are small, but needs to be corrected in the chiral broken phase.
A calculation with a fully dressed quark loop is work in progress.

Finally we have to specify our choice of the quark-gluon vertex, which appears in
the quark self-energy and in the quark loop. We use the same construction
as in \cite{Fischer:2010fx}. It is given by

\bea
\Gamma_\mu(p,k;q) &=& \gamma_\mu\cdot\Gamma(p^2,k^2,q^2) \cdot 
\left(\delta_{\mu,4}\frac{C(p)+C(q)}{2} + \delta_{\mu,i}\frac{A(p)+A(q)}{2} \right), \\
\Gamma(p^2,k^2,q^2) &=& \frac{d_1}{d_2+q^2} + \!\frac{q^2}{\Lambda^2+q^2}
\left(\frac{\beta_0 \alpha(\mu)\ln[q^2/\Lambda^2+1]}{4\pi}\right)^{2\delta}.
\eea
In the UV resummed perturbation theory is recovered with
$\beta_0 = (11N_c-2N_f)/3$,
$\alpha_\mu = 0.3$,
$\delta = -9N_c/(44N_c - 8N_f)$ and the scale
$\Lambda = 1.4$ GeV, which is inherited from the lattice gluon.
In the IR we have two parameters, $d_1$ and $d_2$, which have been fixed to
$d_2 = 0.5$ GeV$^2$ and
$d_1 = 7.6$ GeV$^2$ for $SU(2)$ and $d_1 = 4.6$ GeV$^2$ for $SU(3)$.

\section{Results}
\subsection{Quenched QCD}
\begin{figure}[b]
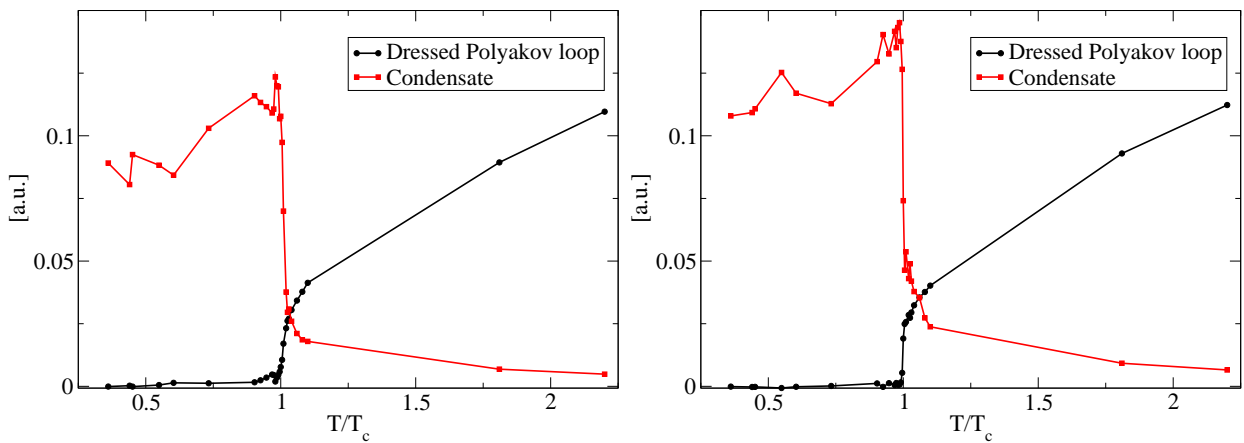

\includegraphics[width=0.44\textwidth]{quenchedNewDataSU2}
\includegraphics[width=0.44\textwidth]{quenchedNewData}
\caption{Dressed Polyakov loop and quark condensate for $SU(2)$ (left) and $SU(3)$ (right).} \label{fig:quenched}
\end{figure}
As already mentioned above, the phase transition of quenched QCD (i.e. without
the quark-loop contributions in the Yang-Mills sector) can be determined from the
quark DSE using quenched lattice data for the temperature dependent gluon propagator 
as input. Of course, the quality of the results then depends on the statistic and 
systematic error of the lattice. Starting from the pioneering work of 
Ref.~\cite{Cucchieri:2007ta}, these have been improved in \cite{Fischer:2010fx} and
analyzed in more detail in recent works 
\cite{Bornyakov:2011jm,Cucchieri:2011di,Aouane:2011fv,MPSS}. In general, however, 
it seems fair to say that in particular systematic errors due to volume and 
discretization artifacts at small momenta are not yet well under control. This is
particularly true in the vicinity of the critical temperature. Consequently, it
proved difficult to distinguish the order of the phase transition between the 
two-color and three-color cases investigated in Ref.~\cite{Fischer:2010fx}. Here
we present updated and improved results for the quark condensate and the dressed
Polyakov loop using the high-statistics lattice data of Ref.~\cite{MPSS} as input, 
which are also carried out on a much finer temperature grid than the ones 
of \cite{Fischer:2010fx}.

Fig.~\ref{fig:quenched} shows how the dressed Polyakov loop and the quark 
condensate change with temperature in the cases of two and three colors. For 
the normalization of the temperature scale we use the transition temperatures which
have been determined from the Polyakov loop on the lattice and compare with
our results obtained from the quark DSE. Indeed, both order parameters show a 
rapid change at $T_c$, signaling the (approximate) restoration of chiral symmetry 
and breaking of centre symmetry at the very same temperature in agreement with the 
critical temperature determined from the lattice. With the finer temperature grid 
and the better statistics compared to \cite{Fischer:2010fx}, the behavior of the 
dressed Polyakov loop is now clearly distinguishable between the $SU(2)$ and $SU(3)$ 
cases, pointing towards a second order phase transition 
for $SU(2)$ and a weak first order for $SU(3)$, again in agreement with the
expectations. The situation is less clear for the chiral transition, although also
here we observe a steeper fall for the $SU(3)$-case. The behavior of the quark 
condensate for temperatures below the critical one, i.e. the rise with temperature
combined with the sharp drop at $T_c$ has also been seen in quenched lattice
calculations \cite{Buividovich:2008ip}. Nevertheless, it may very well be that the 
quantitative aspects of this rise are subject to the systematic uncertainties of the 
lattice gluon data \cite{Cucchieri:2011di}. These uncertainties are also reflected 
in the 'noisy' behavior of the quark condensate. Nevertheless it is remarkable that 
below $T_c$ the dressed Polyakov loop is consistent with zero, signaling conserved 
centre symmetry. 

\subsection{Unquenched QCD at finite temperature and chemical potential}

\begin{figure}
\centering
\includegraphics[width=0.40\textwidth]{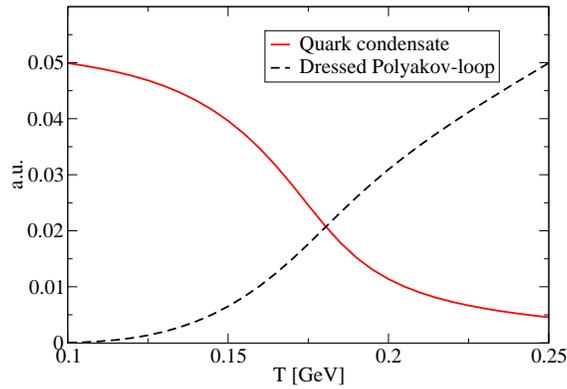}
\caption{Dressed Polyakov loop and quark condensate in two flavor QCD as 
function of temperature at zero chemical potential.} \label{fig:condAndSigNf2}
\end{figure}

We now include two flavors of quarks via the quark loop as explained above.
The effect of the matter sector on the gluon is a reduction of the dressing functions,
which leads to a reduced interaction strength in the quark self-energy, and therefore
to a smaller critical temperature.
In Fig.~\ref{fig:condAndSigNf2} we show the evolution of the order parameters at $\mu=0$.
What we find is a crossover for both the condensate and the dressed Polyakov loop.
The value for the pseudo-critical temperature is $T_c^{N_f=2}=180\pm 5$ MeV from
the quark condensate and $T_c^{N_f=2}=195\pm 5$ MeV from the dressed Polyakov loop.
The difference in these numbers can be attributed to the crossover nature of the transition.

\begin{figure}
\centering
\includegraphics[width=0.44\textwidth]{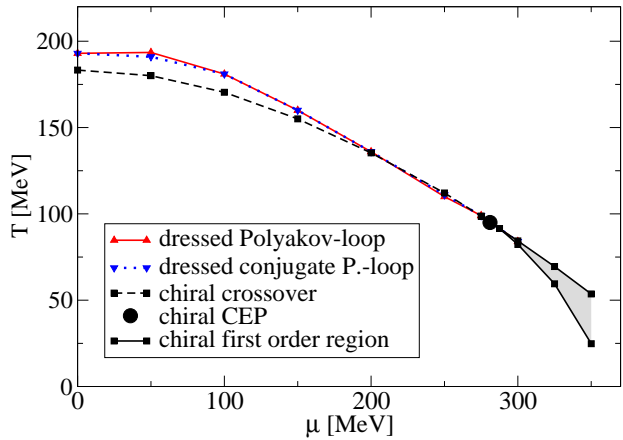}
\caption{The phase boundary for chiral symmetry and confinement
at real chemical potential. The solid lines above the CEP denote
the spinodals which mark the area of coexistence of chiral symmetric and broken solutions
of the DSE.} \label{fig:phaseDiag}
\end{figure}

When we go to $\mu>0$ the condensate for neither periodic ($\varphi=0$) nor anti-periodic ($\varphi=\pi$) boundary conditions
becomes complex. This leads to a difference in $\Sigma_{+1}$ and $\Sigma_{-1}$,
i.e. in the dressed Polyakov loop and its conjugate.
In Fig.~\ref{fig:phaseDiag} the resulting phase diagram of two flavor QCD is shown.
For the chiral transition we observe a crossover up to relatively large values of the chemical
potential where we find a critical endpoint at $(T_{EP},\mu_{EP}) \approx (95,280)$ MeV.
Since $\mu_{EP}/T_{EP}\approx 3 \gg 1$, this result suggests that the CEP is outside the
reach of lattice QCD.
For the confinement/deconfinement transition we observe that the critical temperature extracted
from the dressed Polyakov loop and its conjugate is nearly equal, and close to that extracted
from the quark condensate. As the chemical potential is increased the crossover becomes steeper
and the two transition lines come closer together, meeting at around $\mu \approx 200$ MeV.

Both results, the CEP at large $\mu$ and the coinciding phase transitions agree well with
results from the PQM model \cite{Herbst:2010rf} beyond mean field, where the matter
back-reaction on the Yang-Mills sector is also taken into account.

We should note here that at the chemical potentials where we find the CEP our truncation
scheme becomes less reliable, since the influence of Baryons is neglected.
It may therefore be advised to rephrase our results as an exclusion of the CEP
in the $\mu/T<1$ region. This is consistent with longstanding predictions from 
investigations of the curvature of the chiral critical surface in the Columbia 
plot \cite{de Forcrand:2002ci} and also with recent lattice results on the 
curvature of the chiral and deconfinement crossover lines at small chemical 
potential \cite{Endrodi:2011gv}.

\section{Conclusion}

We have presented a truncation scheme for the Dyson-Schwinger equations of QCD where
we take data from a lattice calculation for the temperature dependent quenched gluon,
and introduce the quark loop for studies of unquenched QCD, namely at finite chemical 
potential.

Within this truncation we investigated the behavior of the quark condensate as
an order parameter for chiral symmetry breaking, and of the dressed Polyakov loop
as an order parameter for confinement.
In the quenched case at $\mu=0$ we found that the order parameters reproduce
the lattice input, hinting at a second order phase transition for $SU(2)$
and a first order phase transition for $SU(3)$.
At finite density we found that thermal fluctuations from the matter sector lead to
a critical endpoint at large densities while chiral and deconfinement transitions 
coincide. Our results serve as a basis for further studies of hot and dense QCD.

\section{Acknowledgements}
We are grateful to Jens A. Mueller for collaboration on part of the 
results summarized here. 
This work has been supported by the Helmholtz Young 
Investigator Grant VH-NG-332 and the Helmholtz International Center 
for FAIR within the LOEWE program of the State of Hesse.


\begin{thebibliography}{99}

%\cite{Fukushima:2003fw}
\bibitem{Fukushima:2003fw}
  K.~Fukushima,
  %``Chiral effective model with the Polyakov loop,''
  Phys.\ Lett.\  {\bf B591 } (2004)  277-284.
  %[hep-ph/0310121].
%\cite{Ratti:2005jh}
%\bibitem{Ratti:2005jh}
  C.~Ratti, M.~A.~Thaler, W.~Weise,
  %``Phases of QCD: Lattice thermodynamics and a field theoretical model,''
  Phys.\ Rev.\  {\bf D73 } (2006)  014019.
  %[hep-ph/0506234].

%\cite{Schaefer:2007pw}
\bibitem{Schaefer:2007pw}
  B.~-J.~Schaefer, J.~M.~Pawlowski, J.~Wambach,
  %``The Phase Structure of the Polyakov--Quark-Meson Model,''
  Phys.\ Rev.\  {\bf D76 } (2007)  074023.
  %[arXiv:0704.3234 [hep-ph]].

%\cite{Herbst:2010rf}
\bibitem{Herbst:2010rf}
  T.~K.~Herbst, J.~M.~Pawlowski, B.~-J.~Schaefer,
  %``The phase structure of the Polyakov--quark-meson model beyond mean field,''
  Phys.\ Lett.\  {\bf B696 } (2011)  58-67.
  %[arXiv:1008.0081 [hep-ph]].

%\cite{Skokov:2010uh}
\bibitem{Skokov:2010uh}
  V.~Skokov, B.~Friman, K.~Redlich,
  %``Quark number fluctuations in the Polyakov loop-extended quark-meson model at finite baryon density,''
  Phys.\ Rev.\  {\bf C83 } (2011)  054904.
  %[arXiv:1008.4570 [hep-ph]];  
%\cite{Skokov:2010wb}
%\bibitem{Skokov:2010wb}
  V.~Skokov, B.~Stokic, B.~Friman, K.~Redlich,
  %``Meson fluctuations and thermodynamics of the Polyakov loop extended quark-meson model,''
  Phys.\ Rev.\  {\bf C82 } (2010)  015206.
  %[arXiv:1004.2665 [hep-ph]].

%\cite{Kojo:2009ha}
\bibitem{Kojo:2009ha}
  T.~Kojo, Y.~Hidaka, L.~McLerran {\it et al.},
  %``Quarkyonic Chiral Spirals,''
  Nucl.\ Phys.\  {\bf A843 } (2010)  37-58.
  %[arXiv:0912.3800 [hep-ph]].

%\cite{Nickel:2009ke}
\bibitem{Nickel:2009ke}
  D.~Nickel,
  %``How many phases meet at the chiral critical point?,''
  Phys.\ Rev.\ Lett.\  {\bf 103 } (2009)  072301.
  %[arXiv:0902.1778 [hep-ph]].

%\cite{McLerran:2007qj}
\bibitem{McLerran:2007qj}
  L.~McLerran, R.~D.~Pisarski,
  %``Phases of cold, dense quarks at large N(c),''
  Nucl.\ Phys.\  {\bf A796 } (2007)  83-100.
  %[arXiv:0706.2191 [hep-ph]].

%\cite{Fischer:2009wc}
\bibitem{Fischer:2009wc}
  C.~S.~Fischer,
  %``Deconfinement phase transition and the quark condensate,''
  Phys.\ Rev.\ Lett.\  {\bf 103 } (2009)  052003;
  %[arXiv:0904.2700 [hep-ph]].
%\cite{Fischer:2009gk}
%\bibitem{Fischer:2009gk}
  C.~S.~Fischer, J.~A.~Mueller,
  %``Chiral and deconfinement transition from Dyson-Schwinger equations,''
  Phys.\ Rev.\  {\bf D80 } (2009)  074029.
  %[arXiv:0908.0007 [hep-ph]].

%\cite{Fischer:2010fx}
\bibitem{Fischer:2010fx}
  C.~S.~Fischer, A.~Maas and J.~A.~Muller,
  %``Chiral and deconfinement transition from correlation functions: SU(2) vs.
  %SU(3),''
  Eur.\ Phys.\ J.\  C {\bf 68} (2010) 165.
  %[arXiv:1003.1960 [hep-ph]].

%\cite{Fischer:2011mz}
\bibitem{Fischer:2011mz}
  C.~S.~Fischer, J.~Luecker, J.~A.~Mueller,
  %``Chiral and deconfinement phase transitions of two-flavour QCD at finite temperature and chemical potential,''
  Phys.\ Lett.\  {\bf B702 } (2011)  438-441.
  %[arXiv:1104.1564 [hep-ph]].

%\cite{Braun:2007bx}
\bibitem{Braun:2007bx}
  J.~Braun, H.~Gies, J.~M.~Pawlowski,
  %``Quark Confinement from Color Confinement,''
  Phys.\ Lett.\  {\bf B684 } (2010)  262-267.
  %[arXiv:0708.2413 [hep-th]].

%\cite{Braun:2009gm}
\bibitem{Braun:2009gm}
  J.~Braun, L.~M.~Haas, F.~Marhauser, J. M. Pawlowski,
  %``On the relation of quark confinement and chiral symmetry breaking,''
  Phys.\ Rev.\ Lett.\  {\bf 106 } (2011)  022002.



%\cite{Edwards:2010ew}
\bibitem{Edwards:2010ew}
  S.~R.~Edwards, A.~Sternbeck, L.~von Smekal,
  %``Exploring a hidden symmetry with electrically charged quarks,''
  PoS {\bf LATTICE2010 } (2010)  275.
  %[arXiv:1012.0768 [hep-lat]].

%\cite{Fischer:2011pk}
\bibitem{Fischer:2011pk}
  C.~S.~Fischer, J.~A.~Mueller,
  %``On critical scaling at the QCD N_f=2 chiral phase transition,''
  Phys.\ Rev.\  {\bf D84 } (2011)  054013.
  %[arXiv:1106.2700 [hep-ph]].



%\cite{Gattringer:2006ci}
\bibitem{Gattringer:2006ci}
  C.~Gattringer,
  %``Linking confinement to spectral properties of the Dirac operator,''
  Phys.\ Rev.\ Lett.\  {\bf 97} (2006) 032003.
  %[arXiv:hep-lat/0605018].
  %%CITATION = PRLTA,97,032003;%%

%\cite{Synatschke:2007bz}
\bibitem{Synatschke:2007bz}
  F.~Synatschke, A.~Wipf and C.~Wozar,
  %``Spectral sums of the Dirac-Wilson operator and their relation to the
  %Polyakov loop,''
  Phys.\ Rev.\  D {\bf 75} (2007) 114003;
  %[arXiv:hep-lat/0703018].
  %%CITATION = PHRVA,D75,114003;%%

%\cite{Bilgici:2008qy}
\bibitem{Bilgici:2008qy}
  E.~Bilgici, F.~Bruckmann, C.~Gattringer and C.~Hagen,
  %``Dual quark condensate and dressed Polyakov loops,''
  Phys.\ Rev.\  D {\bf 77} (2008) 094007.
  %[arXiv:0801.4051].
  %%CITATION = PHRVA,D77,094007;%%

%\cite{Cucchieri:2007ta}
\bibitem{Cucchieri:2007ta}
  A.~Cucchieri, A.~Maas, T.~Mendes,
  %``Infrared properties of propagators in Landau-gauge pure Yang-Mills theory at finite temperature,''
  Phys.\ Rev.\  {\bf D75 } (2007)  076003.
  %[hep-lat/0702022].

%\cite{Bornyakov:2011jm}
\bibitem{Bornyakov:2011jm}
  V.~G.~Bornyakov, V.~K.~Mitrjushkin,
  %``Lattice QCD gluon propagators near transition temperature,''
  [arXiv:1103.0442 [hep-lat]].  

%\cite{Cucchieri:2011di}
\bibitem{Cucchieri:2011di}
  A.~Cucchieri, T.~Mendes,
  %``Electric and magnetic Landau-gauge gluon propagators in finite-temperature SU(2) gauge theory,''
  PoS {\bf FACESQCD } (2010)  007.
  %[arXiv:1105.0176 [hep-lat]].

%\cite{Aouane:2011fv}
\bibitem{Aouane:2011fv}
  R.~Aouane, V.~Bornyakov, E.~-M.~Ilgenfritz, V.~Mitrjushkin, M.~Muller-Preussker, A.~Sternbeck,
  %``Landau gauge gluon and ghost propagators at finite temperature from quenched lattice QCD,''
  [arXiv:1108.1735 [hep-lat]].  

\bibitem{MPSS} A.~Maas, J.~M.~Pawlowski, L.~v.~Smekal, D.~Spielmann, [arXiv:1110.6340 [hep-lat]].

%\cite{Buividovich:2008ip}
\bibitem{Buividovich:2008ip}
  P.~V.~Buividovich, E.~V.~Luschevskaya, M.~I.~Polikarpov,
  %``Finite-temperature chiral condensate and low-lying Dirac eigenvalues in quenched SU(2) lattice gauge theory,''
  Phys.\ Rev.\  {\bf D78 } (2008)  074505.
  %[arXiv:0809.3075 [hep-lat]].

%\cite{de Forcrand:2002ci}
\bibitem{de Forcrand:2002ci}
  P.~de Forcrand, O.~Philipsen,
  %``The QCD phase diagram for small densities from imaginary chemical potential,''
  Nucl.\ Phys.\  {\bf B642 } (2002)  290-306;
  %[hep-lat/0205016]. 
  %\cite{deForcrand:2008zi}
%\bibitem{deForcrand:2008zi}
%  P.~de Forcrand, O.~Philipsen,
  %``The curvature of the critical surface (m(u,d),m(s))**crit(mu): A Progress report,''
  PoS {\bf LATTICE2008 } (2008)  208.
  %[arXiv:0811.3858 [hep-lat]].

%\cite{Endrodi:2011gv}
\bibitem{Endrodi:2011gv}
  G.~Endrodi, Z.~Fodor, S.~D.~Katz, K.~K.~Szabo,
  %``The QCD phase diagram at nonzero quark density,''
  JHEP {\bf 1104 } (2011)  001.
  %[arXiv:1102.1356 [hep-lat]].
  
\end{thebibliography}
\end{document}